\documentclass[prl,preprint]{revtex4}

\usepackage{epsfig,graphics}
\usepackage[cmex10]{amsmath}

\begin{document}
\title{Optical Detection of Single Nanoparticles with a Sub-wavelength Fiber-Taper}
\author{Jiangang Zhu, Sahin Kaya Ozdemir, and Lan Yang}
\affiliation{Department of Electrial and Systems Engineering, Washington University, St. Louis, Missouri, 63130 USA}

\begin{abstract}
A nanoparticle detection scheme with single particle resolution is presented. The sensor contains only a taper fiber thus offering the advantages of compactness and installation flexibility. Sensing method is based on monitoring the transmitted light power which shows abrupt jumps with each particle binding to the taper surface. The experimental validation of the sensor is demonstrated with polystyrene nanoparticles of radii 120 nm and 175 nm in the 1550 nm wavelength band.
\end{abstract}

\maketitle

\section{Introduction}
Nanoparticle detection and characterization techniques have been widely sought as the awareness of the potential benefits and risks of the continuously generated by-product or massively synthesized nanoparticles are increasing. Nanoparticles of special interests range from biological agents and virions to specially synthesized semiconductor, metal and polymer nanoparticles. While the detection of the former is important for biodefense and early detection of pandemic outbreaks, detection and characterization of the latter group of nanoparticles are indispensable for their broad range of applications in nanotechnology.

In addition to conventional microscopic techniques which, despite their high sensitivity and resolution, are not suitable for field measurements due to their expensive and bulky constructions, long processing times and the necessity of pre-treatment (labeling with fluorescent dyes, etc) of the particles, there exist many variations of optical particle counters \cite{opc1,opc2} which rely on light scattering measurements, allow field measurements, detect and count individual or ensemble of particles. These counters require off-axis detectors for the collection of the scattered light, have bulky configurations and require sophisticated signal processing.

Recently, there is a growing interest for nanoparticle detection using nano- and micro-scale sensors, which, with their unprecedented sensitivity, have the potential for in-situ sensing. The nano/micro-mechanical sensors \cite{Weighing,NEMS} detect particles by monitoring the resonance frequency changes caused by the additional effective mass of the binding particles, while resonator-based micro/nano-optical resonator sensors rely on either the resonance frequency shift \cite{Arnold,vollmer,Fan} or mode splitting \cite{Jiangang1} due to the change in the effective polarizability of the resonator-environment system upon particle binding. The latter have shown to detect and count individual nanoparticles as small as radius $R$=30 nm. This high sensitivity is attributed to the resonance enhanced interaction between the particle and the evanescent tail of the light field due to tight light confinement and extended interaction time provided by the resonator. These sensors require a fiber taper to couple the light into and out of the resonator from a tunable laser, whose wavelength is continuously scanned to monitor the changes in the resonance modes thus making these highly compact and sensitive sensors expensive.

In the past few decades, much effort has been devoted to developing fiber optic evanescent field-based sensors using tapered optical fibers. These sensors have been shown to be very sensitive to changes of the surrounding medium providing a compact, inexpensive and in-line sensing platform for measuring physical parameters such as temperature and refractive index, as well as for quantitatively detecting the concentration of chemical compounds \cite{taper1,taper2}.

In this Letter, we demonstrate that a tapered optical fiber with sub-micron diameter can be effectively used for label-free detection and counting of single nanoparticles as each individual nanoparticle enters the evanescent field of the nano-taper. The proposed scheme eliminates the need for tunable lasers, bulky optical components and lengthy signal processing tasks \cite{NEMS, vollmer, Jiangang1}. It enables a versatile yet practical, portable, compact and low-cost single nanoparticle detection platform with comparable sensitivity to existing schemes.

\section{Principle}

The underlying principle of the proposed sensing mechanism is based on optical scattering of the evanescent field of a tapered optical fiber when a sub-wavelength (Rayleigh) scatterer enters the mode volume. In a single mode fiber, the light propagates as a core mode, i.e., most of the energy is confined within the core. However, as the fiber is tapered down, the core area becomes smaller and light spreads out into the cladding and consequently the core mode adiabatically transforms into a cladding mode leading to a highly-confined field at the cladding-medium interface with an evanescent portion in the surrounding medium. This cladding mode is then adiabatically converted back to the propagating core mode after the tapered region. Thus, the tapered region facilitates access to the evanescent field, allowing it to interact with the surrounding medium. Subsequently, making the tapered region susceptible to any perturbations (e.g., changes in refractive index, temperature, humidity, absorbtion, scattering, etc) in the medium. When a sub-wavelength spherical particle of radius $R$ and permittivity $\varepsilon_p$ is placed in the evanescent field ${\bf E}_0$ of the tapered region, it will induce a scattering loss which can be described by the field of an induced dipole moment ${\bf p}=\alpha\varepsilon_{\rm m} {\bf E}_0$ where $\alpha=4\pi R^3(\varepsilon_{\rm s}-\varepsilon_{\rm p})/(\varepsilon_p+2\varepsilon_{\rm m})$ is the polarizability of the scatterer and $\varepsilon_{\rm m}$ is the permittivity of the surrounding medium. This scattering loss to the environment will then lead to decrease in the transmitted power at the output port of the fiber. Since the polarizability $\alpha$ is a function of the shape and the size of the particle, and the permittivity (i.e, refractive index) contrast of the particle and the surrounding medium, the loss in the transmission should contain information on these properties of the particle. Thus, monitoring the changes in the transmission, one will be able to detect and quantify the polarizability of the particles entering the evanescent field.

\begin{figure}[!t]
\centering
\includegraphics[width=2.7in]{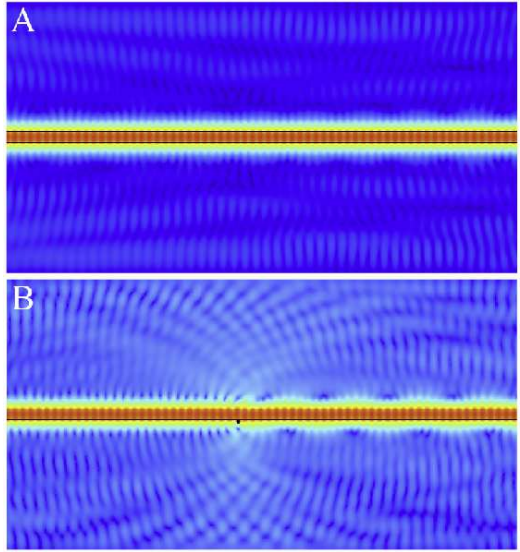}
\caption{Results of numerical simulations performed using COMSOL showing the pattern of the electrical field of the light scattered by a nanoparticle adsorbed on a nano-taper. (a) Without particle. (b) With particle. The nano-taper has a thickness of 0.8 $\mu m$ and refractive index of 1.45, light wavelength is 1550 nm and the nanoparticle has refractive index of 1.59 and a radius of 150 nm.}
\label{fig1}
\end{figure}

Numerical simulations of electric field around a taper in the presence of a subwavelength nanoparticle clearly show the particle-induced disturbance and scattering (Fig. \ref{fig1}).

\section{Experiment}

\begin{figure}[h]
\centering
\includegraphics[width=3in]{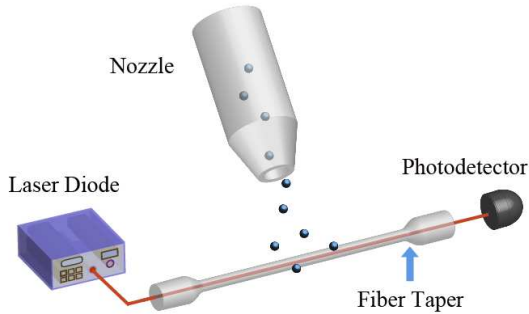}
\caption{Experimental setup showing single particles deposited onto the the tapered fiber.}
\label{fig2}
\end{figure}

A schematics of the set-up used in our experiments is given in Fig.\ref{fig2}. The tapered fibers were prepared by heating and pulling a standard communication single-mode fiber ($R_{\rm core}$=4 $\mu$m and $R_{\rm clad}$=62.5 $\mu m$) above a hydrogen flame. The thickness and length of the waist of the fabricated taper was estimated to be 0.8 ${\rm \mu m}$ and 3 ${\rm mm}$, respectively. Light was provided by an unmodulated continuous wave (CW) laser diode with fixed power around $P$=2 ${\rm mW}$ at wavelength of $\lambda$=1.55 ${\rm \mu m}$. The transmitted light power was measured with a photodetector (PD; bandwidth: 125 ${\rm MHz}$) whose output was then acquired to a computer. Polystyrene (PS) nanoparticles of refractive index $n_{\rm s}=\sqrt{\varepsilon_{\rm s}}=1.59$ of mean radii $R$=120 ${\rm nm}$ and $R$=175 ${\rm nm}$ were used to test the performance of the nano-taper sensor for detecting nanoparticles at single particle resolution.

Polystyrene nanoparticles were deposited onto the taper using a set-up which consists of an atomizer, a differential mobility analyzer (DMA) and a nozzle with an inner tip diameter of 80 ${\rm \mu m}$ \cite{Jiangang1}.  The particles are carried out by compressed air using a collision atomizer and then neutralized by a radioactive source such that they have a narrow charge distribution. The DMA classifies particles according to their electrical mobility, resulting in a narrow size distribution. The filted particles exit through the output slit and are subsequently channeled to the taper waist with a micro-nozzle.

\begin{figure}[!t]
\centering
\includegraphics[width=3.2in]{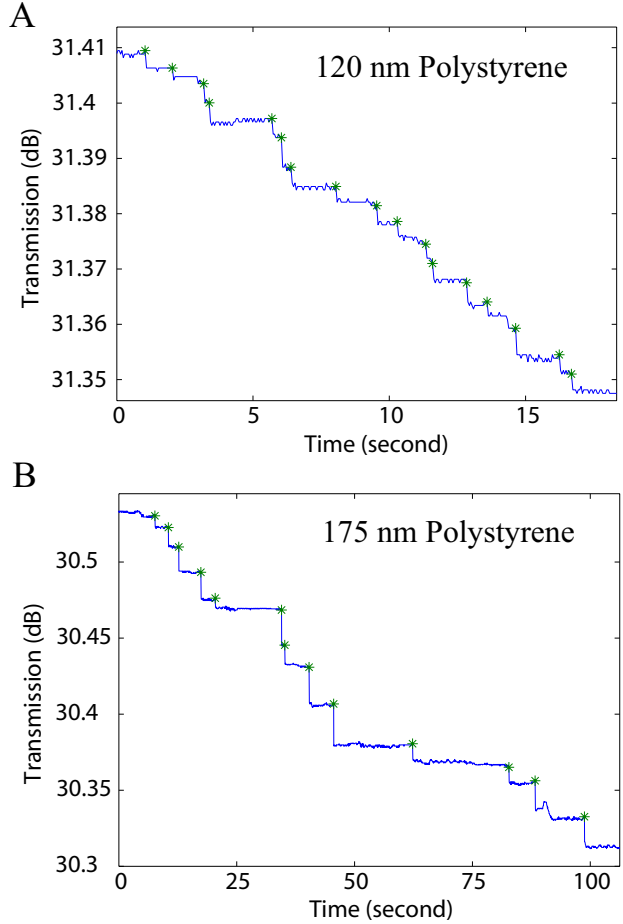}
\caption{The change in the transmission as polystyrene nanoparticles bind to a nano-taper. (a) $R$=120$\pm3$ ${\rm nm}$, and (b) $R$=175$\pm4$ ${\rm nm}$. The discrete jump heights larger than the amount of power fluctuations are considered to be induced by particle binding and labeled with '*'. The data capturing rate is 20 points per second.}
\label{fig3}
\end{figure}

Figure \ref{fig3} shows the changes in the transmission as a function of time as PS nanoparticles are deposited onto the taper waist section. Each discrete downward jump in Fig. \ref{fig3} signals the binding of a single nanoparticle. Thus, by counting the number of these jumps, one can count the number of particles entering the field of the taper. Height of the jumps, which reflect the effective scattering loss, varies with the position of the particles along the taper waist as well as their distance from each other. The first is attributed to the slight non-uniformity of the waist diameter which leads to varying local field intensities along the taper. The latter is due to multi-particle scattering and the modification of the local field distribution due to deposited particles. It should be noted that the particles falling out of tapered region do not interact with the evanescent field and thus cannot be detected.

\begin{figure}[h]
\centering
\includegraphics[width=3.2in]{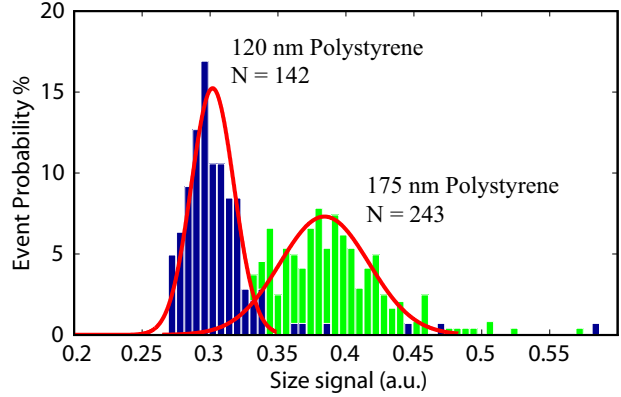}
\caption{Measured size signal ($h^{1/6}$) distributions for polystyrene nanoparticles of $R$=120$\pm3$ ${\rm nm}$ (left peak) and $R$=175$\pm4$ ${\rm nm}$ (right peak). $N$ denotes the number of particles detected during the experiments.}
\label{fig4}
\end{figure}

The scattering cross-section and hence the effective scattering loss induced by a sub-wavelength Rayleigh scatterer is proportional to $\alpha^2$, or $R^6$. Thus, the heights of the jumps in the power transmission carry the information on the particle cross-section or the particle size. To confirm this, we use $h^{1/6}$ as the size signal where $h$ denotes the height of each discrete jump in the transmission signal. Figure \ref{fig4} shows the recorded distribution of $h^{1/6}$ measured for PS particles of two different sizes. The clear separation of the peaks and the small overlap between the tails of the distributions suggest that the two sizes of particles are well resolved. The standard deviations of the distributions are larger than those provided by the manufacturers mainly due to multi-particle effects and non-uniformity of the taper waist. The laser power noise and detector noise also contribute to this deviation, especially for the $R$=120 nm case, when particle induced signals are closer to the noise level.


\section{Conclusion}

We demonstrated real-time detection and counting of single PS nanoparticles using a tapered-fiber with sub-micron taper waist. The proposed scheme does not need tunable lasers and offers single-particle resolution, ease of fabrication, low-cost and versatility. Thus, it provides an alternative and competitive platform to existing technologies with comparable sensitivities. We should note here that since the scattering loss scales as $\lambda^{-4}$, and taper field cross-section scales as $\lambda^{2}$, the sensitivity of the scheme for detecting particles with much smaller size are within reach by replacing the infrared laser source with shorter wavelength lasers, i.e., visible wavelengths. Moreover, a number of tapered fibers can be fixed together to form an array which will increase the sensing area thus improving particle capturing efficiency.

This detection scheme can be easily applied to planar waveguide structures. In aqueous medium, our method combined with recently developed waveguide particle trapping and transporting techniques \cite{trapping, Richardson1, transport, Richardson2} can form a complete nanoparticle detection and sorting platform on a chip. This technology is not limited to the detection of dielectric nanoparticles, but it can be used for the detection of metal nanoparticles and large bio-particles, as well. The selectivity can be achieved by applying recognition coatings on the tapered region.

\section*{Acknowledgment}
This work was supported by the National Science Foundation under Grant No. 0954941. The authors thank L. He, M. Faraz, and W. Kim for discussions and Prof. D.-R. Chen for providing DMA.

\end{document}